\documentclass[showpacs,amsmath,amssymb,onecolumn,prx,nofootinbib]{revtex4-1}

\usepackage[dvips]{graphicx} 
\usepackage{amsfonts}
\usepackage{amssymb}
\usepackage{amscd}
\usepackage{amsmath}    
\usepackage{enumerate}
\usepackage{epsfig}
\usepackage{subfigure}
\usepackage{xcolor}
\usepackage{amsthm}

\newcommand{\bra}[1]{\mbox{$\left\langle #1 \right|$}}
\newcommand{\ket}[1]{\mbox{$\left| #1 \right\rangle$}}

\begin{document}

\title{Security of modified Ping-Pong protocol in noisy and lossy channel}

\author{Yun-Guang Han$^1$$^,$$^2$$^\#$, Zhen-Qiang Yin$^1$$^,$$^2$$^\#$, Hong-Wei Li$^1$$^,$$^2$, Wei Chen$^1$$^,$$^2$$^{*}$, Shuang Wang$^1$$^,$$^2$$^{*}$, Guang-Can Guo$^1$$^,$$^2$, Zheng-Fu Han$^1$$^,$$^2$$^{*}$\\}

\affiliation{$^1$ Key Laboratory of Quantum Information,University of Science and Technology of China,Hefei 230026,China\\
$^2$ Synergetic Innovation Center of Quantum Information $\&$ Quantum Physics, University of Science and Technology of China, Hefei, Anhui 230026, China\\
$^\#$ Both authors contributed equally to this work.\\
$^{*}$ Correspondence to: (W.C.) kooky@mail.ustc.edu.cn,\\(S.W.) wshuang@ustc.edu.cn, \\ (Z.F.H) zfhan@ustc.edu.cn.
}

\begin{abstract}
The ``Ping-Pong" (PP) protocol is a two-way quantum key protocol based on entanglement. In this protocol, Bob prepares one maximally entangled pair of qubits, and sends one qubit to Alice. Then, Alice performs some necessary operations on this qubit and sends it back to Bob. Although this protocol was proposed in 2002, its security in the noisy and lossy channel has not been proven. In this report, we add a simple and experimentally feasible modification to the original PP protocol, and prove the security of this modified PP protocol against collective attacks when the noisy and lossy channel is taken into account. Simulation results show that our protocol is practical.

\end{abstract}


\maketitle

\section*{Introduction}

Quantum key distribution (QKD) \cite{BB84, RMP2002} allows two remote parties (Alice and Bob) to establish unconditional secure-key bits. Research of QKD mainly concerns one-way protocols. BB84 \cite{BB84} protocol is the most commonly used one-way QKD protocol. One-way QKD protocols need the information carriers to be transferred from Alice to Bob via the quantum channel to generate the secure-key bits. For example, in BB84, Alice randomly prepares her qubits into one of the quantum states $|0\rangle$, $|1\rangle$, $|+\rangle=(|0\rangle+|1\rangle)/\sqrt{2}$, and $|-\rangle=(|0\rangle-|1\rangle)/\sqrt{2}$, and sends them to Bob via an untrusted quantum channel. Then Bob performs some measurements on the incoming qubits and key bits are generated.

In the past decade, some two-way QKD protocols have been proposed \cite{PP2002,Cai2004,Cai_CPL, LM2005, N09} and experimentally realized\cite{Exp2008,Exp2012,experiment1 of N09, experiment2 of N09, experiment3 of N09}. In these protocols, Bob prepares some quantum states and sends them to Alice, Alice performs encoding operations on the received states and sends them back in the same quantum channel, then Bob makes some measurements and gets the key bits. Typical examples of these protocols are ``Ping-Pong'' (PP) \cite{PP2002} protocol, LM05 \cite{LM2005} protocol and N09 \cite{N09} protocol.  The first one uses entanglement while the last two protocols not. These protocols are all deterministic, which means Bob can obtain Alice's key bits directly, without a basis reconciliation step.

The major difficulty of the security proof for the two-way QKD protocols is that the eavesdropper, Eve, may attack the travel qubit on both the forward (Bob-Alice) channel and the backward (Alice-Bob) channel. This makes the security analysis quite complicated. Fortunately, for the LM05 protocol, its security has been proven in different ways \cite{Cai2011,Cai2012,Renner2013} recently, and N09 protocol is also proven to be secure in Refs. \cite{security of N09, security2 of N09}. A super dense coding (SDC) protocol similar to PP protocol is also proved secure\cite{Renner2013}. However, these proofs are all based on the assumption that there are no losses in the channel and detectors. In fact, the security of PP protocol has been challenged by channel loss \cite{Cai_PRL,Attack2003,Zhang2004,BF2008}. Hence, the security of PP protocol when loss is considered remains an open question until now.

In this paper, we propose a modified Ping-Pong protocol and prove its security against collective attacks in noisy and lossy channel.  To the best of our knowledge, this is the first security proof of PP protocol considering the loss in channel.

\section*{Original Ping-Pong protocol}

It's beneficial to give a brief description of the PP protocol.
Consider that Bob prepares two-qubits maximally entangled state $\ket{\Phi^+}=(1/\sqrt{2})(\ket{00}+\ket{11})$,
where $\ket{0}$ and $\ket{1}$ are the eigenstates of Pauli matrix $\sigma_z$. He sends one of the qubits (travel qubit) to Alice via an untrusted channel and retains the other (home qubit) in her quantum memory. Then, Alice randomly chooses message mode or control mode to proceed. In message mode, Alice performs an unitary operation $I$ to the incoming qubit to encode classical bit 0, or $\sigma_z$ to encode bit 1. After her encoding operation, Alice sends this qubit back to Bob. When Bob receives the travel qubit, he performs a Bell measurement on both qubits to decode Alice's bit. In control mode, Alice measures the travel qubit in the basis $B_z=\{\ket{0},\ket{1}\}$ and announces her measurement outcome through an authenticated channel. On receiving Alice's measurement result, Bob also measures his home qubit with $B_z$ and compare his own measurement result with Alice's. If Alice's and Bob's measurement results coincide, the PP protocol continues; otherwise, they may terminate the protocol. Hence the control mode is used to guarantee the security, while the message mode is for key distribution.

In next section, the modified PP protocol will be given.

\section*{Modified Ping-Pong protocol}

In the modified PP protocol, we also use the maximal EPR pair $\ket{\Phi^+}=(1/\sqrt{2})(\ket{00}+\ket{11})$. This protocol process contains four steps:
\begin{enumerate}
  \item
  Bob prepares N pairs of entangled states $\ket{\Phi^+}=(1/\sqrt{2})(\ket{00}+\ket{11})$. He sends half of states (travel qubits) to Alice through a noisy and lossy  quantum channel(forward channel), and keeps the other half (home qubits) in his quantum memory.
  \item
  Alice randomly switches to message mode or control mode with probability $c$ and $1-c$ respectively. In message mode, Alice performs one of the four unitary operations  $I_{0},I_{1},Y_{0}$ and $Y_{1}$ to the incoming states, i.e.,

  $I_{0}\{\ket{v},\ket{0},\ket{1}\}=\{\ket{v},\ket{0},\ket{1}\}$,
  $I_{1}\{\ket{v},\ket{0},\ket{1}\}=\{\ket{v},-\ket{0},-\ket{1}\}$,

  $Y_{0}\{\ket{v},\ket{0},\ket{1}\}=\{\ket{v},\ket{0},-\ket{1}\}$,
  $Y_{1}\{\ket{v},\ket{0},\ket{1}\}=\{\ket{v},-\ket{0},\ket{1}\}$.

where $\ket{v}$ means the vacuum state. Due to the existence of vacuum state, $I_{0}$ and $I_{1}$ are no longer identical. So is $Y_{0}$ and $Y_{1}$.  The probabilities of each operation are all 1/4. For operations $I_{0}$ and $I_{1}$, Alice records her classical information as bit 0. For $Y_{0}$ and $Y_{1}$, Alice records bit 1. Then Alice sends the encoded states back to Bob through the backward channel. In control mode, Alice measures the incoming signals with projectors $\{|v\rangle\langle v|,|0\rangle\langle 0|, |1\rangle\langle 1|\}$. Then Alice records her measurement results.
  \item
  Bob also randomly switches to message mode or control mode with probability $c$ and $1-c$ respectively. In message mode, Bob performs a Bell-states measurement on his home qubit and his received travel qubit to decode Alice's information (i.e., when Alice encodes bit 0, Bob may obtain the Bell state $\ket{\Phi^+}=(1/\sqrt{2})(\ket{00}+\ket{11})$; when Alice encodes bit 1, Bob may obtain $\ket{\Phi^-}=(1/\sqrt{2})(\ket{00}-\ket{11})$). In control mode, Bob measures his reserved qubits with projectors $\{|v\rangle\langle v|,|0\rangle\langle 0|, |1\rangle\langle 1|\}$ and records the measurement results.
  \item
  Alice and Bob publicly announces which trials are in message mode, which are in control mode and their measurements results in control mode. Then
  Alice and Bob can share the probabilities $p_{00}$, $p_{01}$, $p_{0v}$, $p_{10}$, $p_{11}$, and $p_{1v}$ (i.e. $p_{0v}$ is the probability Alice receives a vacuum state when the travel qubit is $\ket{0}$, and other probabilities have the similar meanings). These probabilities will be used to bound the eavesdropper Eve's information on key bits. By sacrificing certain bits for error testing, Alice and Bob can also estimate the error rate $e$ for their key bits.
  Alice and Bob then do classical postprocessing error correction (EC) and privacy amplification (PA) to generate secure-key bits.
\end{enumerate}

  The protocol flow is illustrated in Fig. 1.

\section*{Security Proof of Modified Ping-Pong protocol }
\subsection*{Eve's Attack in the Bob-Alice channel}
Eve's most general collective attack in the Bob-Alice channel can be written in the form
\begin{equation}
\begin{aligned}
  U_{AE}\ket{0}_A\ket{E}=\sqrt{p_{0v}}\ket{v}_A\ket{E_{0v}}+\sqrt{p_{00}}\ket{0}_A\ket{E_{00}}+\sqrt{p_{01}}\ket{1}_A\ket{E_{01}}\\
  U_{AE}\ket{1}_A\ket{E}=\sqrt{p_{1v}}\ket{v}_A\ket{E_{1v}}+\sqrt{p_{10}}\ket{0}_A\ket{E_{10}}+\sqrt{p_{11}}\ket{1}_A\ket{E_{11}}
\end{aligned}
\end{equation}
where $p_{0v}$ is the probability Alice receives a vacuum state when the travel qubit is $\ket{0}$, so is $p_{00}, p_{01},p_{1v},p_{10}$ and $p_{11}$. Normalized vectors{\ket{E_{iv}}} and  $\ket{E_{ij}}$ are possible quantum states of Eve's ancilla.

So the joint density matrix of the travel qubit and Eve's ancilla becomes
\begin{equation}
\begin{aligned}
  \rho^{AE}_{Bob-Alice}&=U_{AE}tr_B P\{\ket{\Phi^+_{AB}}\ket{E}\}U^+_{AE}\\
  &=U_{AE}(\frac{1}{2}\ket{0}_A\bra{0}\otimes\ket{E}\bra{E}+\frac{1}{2}\ket{1}_A\bra{1}\otimes\ket{E}\bra{E})U^+_{AE}\\
  &=\frac{1}{2}P\{\sqrt{p_{0v}}\ket{v}_A\ket{E_{0v}}+\sqrt{p_{00}}\ket{0}_A\ket{E_{00}}+\sqrt{p_{01}}\ket{1}_A\ket{E_{01}}\}\\
  &+\frac{1}{2}P\{\sqrt{p_{1v}}\ket{v}_A\ket{E_{1v}}+\sqrt{p_{10}}\ket{0}_A\ket{E_{10}}+\sqrt{p_{11}}\ket{1}_A\ket{E_{11}}\},
\end{aligned}
\end{equation}
in which,$P\{\ket{x}\}=\ket{x}\bra{x}$.

After receiving the forward qubits, in encoding mode, Alice will encode her key bits onto the forward qubit by the operations $I_{0},I_{1},Y_{0}$ and $Y_{1}$ with the same probability 1/4. The operations $I_{0}$ and $I_{1}$ result in the same encoding key bit $0$, $Y_{0}$ and $Y_{1}$ result in bit $1$.  The probabilities that Alice encodes key bit 0 or 1 are still both $1/2$.

Let us at first consider the case where Bob's travel qubit collapses into $\ket{0}$ (i.e., this case corresponds to the third row of the Eq. (2).). After Alice encoding bit 0, the joint state of Alice and Eve becomes
\begin{equation}
\begin{aligned}
   \rho^{AE0}_{Bob-Alice}&=\frac{1}{2}P\{\sqrt{p_{0v}}\ket{v}_A\ket{E_{0v}}+\sqrt{p_{00}}\ket{0}_A\ket{E_{00}}+\sqrt{p_{01}}\ket{1}_A\ket{E_{01}}\}\\
   &+\frac{1}{2}P\{\sqrt{p_{0v}}\ket{v}_A\ket{E_{0v}}-\sqrt{p_{00}}\ket{0}_A\ket{E_{00}}-\sqrt{p_{01}}\ket{1}_A\ket{E_{01}}\}\\
   &=p_{0v}P\{\ket{v}_A\ket{E_{0v}}\}+P\{\sqrt{p_{00}}\ket{0}_A\ket{E_{00}}+\sqrt{p_{01}}\ket{1}_A\ket{E_{01}}\}.
\end{aligned}
\end{equation}

For the case Alice encodes bit 1, the state is
\begin{equation}
  \rho^{AE1}_{Bob-Alice}=p_{0v}P\{\ket{v}_A\ket{E_{0v}}\}+P\{\sqrt{p_{00}}\ket{0}_A\ket{E_{00}}-\sqrt{p_{01}}\ket{1}_A\ket{E_{01}}\}.
\end{equation}

As described in our modified protocol, Eve cannot obtain any information from the vacuum state. So we can exclude the vacuum state from the joint state and renormalize effective encoding density matrices. Define $\eta_{\rightarrow} =p_{00}+p_{01}, p'_{00}=p_{00}/\eta_{\rightarrow} $ and  $p'_{01}=p_{01}/\eta_{\rightarrow}$ (i.e., $\eta_{\rightarrow}$ can be understood as the efficiency for the forward channel, and is estimated directly in experiment.), then the effective encoding matrices in the orthogonal basis $\{\ket{0}_A\ket{E_{00}},\ket{1}_A\ket{E_{01}}\}$ are given by
\begin{equation}
\begin{aligned}
 \rho^{AE0}= \left(
    \begin{array}{cc}
      p'_{00} & \sqrt{p'_{00}p'_{01}} \\
      \sqrt{p'_{00}p'_{01}} & p'_{01} \\
    \end{array}
  \right),
    \rho^{AE1}= \left(
    \begin{array}{cc}
      p'_{00} & -\sqrt{p'_{00}p'_{01}} \\
      -\sqrt{p'_{00}p'_{01}} & p'_{01} \\
    \end{array}
  \right),\\
\end{aligned}
\end{equation}

\begin{equation}
\begin{aligned}
 \rho^{AE}=&\frac{1}{2}\rho^{AE0}+\frac{1}{2}\rho^{AE1}\\
 &=\left(
    \begin{array}{cc}
      p'_{00} & 0 \\
      0 & p'_{01} \\
    \end{array}
  \right).
 \end{aligned}
\end{equation}

Since the density matrix we get is already diagonal, our following calculations can be very simple. With system $AE$, Eve's Von-Neumann entropies on Alice's key bit $A'$, is given by :

\begin{equation}
\begin{aligned}
  S(A'|AE)&=S(\rho^{A'AE})-S(\rho^{AE})\\
  &=S(\frac{1}{2}|0\rangle_{A'}\langle0|\otimes \rho^{AE0}+\frac{1}{2}|1\rangle_{A'}\langle1|\otimes\rho^{AE1})-S(\rho^{AE})\\
  &=H(\frac{1}{2})+\frac{1}{2}S(|0\rangle_{A'}\langle0|\otimes \rho^{AE0})+\frac{1}{2}S(|1\rangle_{A'}\langle1|\otimes\rho^{AE1})-S(\rho^{AE})\\
  &=1-H(p'_{01}),
\end{aligned}
\end{equation}
where $H$ is the Shannon's binary entropy function.

\subsection*{Eve's Attack in the Alice-Bob channel}

Quantum systems $AE$ on the backward channel can be divided into two events: Bob receives the travel qubit and not. We label these two parts as $\rho^{AE}_{received}$ and $\rho^{AE}_{unreceived}$ . As showed in the prior section, the ratio of the two parts are $\eta_{\leftarrow}$ and $1-\eta_{\leftarrow}$ respectively (i.e., $\eta_{\leftarrow}$ is the efficiency of backward channel, and can be estimated by Alice and Bob in experiment.). Our goal is to get the lower bound of $H(A|E)_{received}$ ,which means the conditional entropy of Alice on Eve in the case Bob receives the backward qubits. From definition, we have
\begin{equation}
  S(A'|A E)_{received}=S(\rho^{A'AE}_{received})-S(\rho^{AE}_{received}),
\end{equation}

\begin{equation}
   S(A'|AE)_{unreceived}=S(\rho^{A'AE}_{unreceived})-S(\rho^{AE}_{unreceived}).
\end{equation}

According to the joint entropy theorem, and noticing that the events Bob receives a returning qubit or not are orthogonal, we obtain
\begin{equation}
  S(\rho^{A'AE})=H(\eta_{\leftarrow})+\eta_{\leftarrow} S(\rho^{A'AE}_{received})+(1-\eta_{\leftarrow})S(\rho^{A'AE}_{unreceived})
\end{equation}

\begin{equation}
  S(\rho^{AE})=H(\eta_{\leftarrow})+\eta_{\leftarrow} S(\rho^{AE}_{received})+(1-\eta_{\leftarrow})S(\rho^{AE}_{unreceived}).
\end{equation}

By (10)-(11),
\begin{equation}
  S(A'|AE)=\eta_{\leftarrow} S(A'|AE)_{received}+(1-\eta_{\leftarrow})S(A'|AE)_{unreceived}
\end{equation}

To obtain the lower bound of $S(A'|AE)_{received}$, it is reasonable to assume that Eve has maximal entropy of Alice's key bit $A'$ in the case Bob doesn't receive the backward qubit. And recall Eq. (7), we get
\begin{equation}
   S(A'|AE)_{received}\geq \frac{\eta_{\leftarrow}-H(p'_{01})}{\eta_{\leftarrow}}=1-\frac{H(p'_{01})}{\eta_{\leftarrow}}.
\end{equation}
Combined with the case that Bob's travel qubit collapsed into $|1\rangle$, Eve's total entropy on Alice is given by:

\begin{equation}
   S(A'|AE)_{received}\geq 1-\frac{H(p'_{01})+H(p'_{10})}{2\eta_{\leftarrow}}.
\end{equation}

\subsection*{Secure key rate}
We assume that the raw key bits are unbiased distributed, which means the error rate Alice encodes 0 or 1 are equal. The error bit $e$ is defined as the probability Alice encodes 0 but the BSM gets $\ket{\Phi^{-}}$ or Alice encodes 1 but the BSM gets $\ket{\Phi^{+}}$. By the authenticated communications, Alice and Bob can estimate the error rate $e$ of their raw key bits. Then they will perform error correction and privacy amplification to generate the secure-key bits. The secure-key rate is given by \cite{Winter}:
\begin{equation}
  R\geq S(A'|AE)_{received}-S(A'|B)\geq 1-\frac{H(p'_{01})+H(p'_{10})}{2\eta_{\leftarrow}}-H(e),
\end{equation}
where, $S(A'|B)$ is the conditional entropy for Bob's key bits to Alice's key bits and equals $H(e)$ under the unbiased distribution assumption.

We can find the the secure-key rate for this modified PP protocol is very simple. Eve's information on key bits can be just bounded by the error rates for forward channel $p'_{01}$, $p'_{10}$ and the efficiency of backward channel $\eta_{\leftarrow}$. The physics behind our proof is that our operation $I_0$, $I_1$, $Y_0$, and $Y_1$ will introduce a phase randomization to Eve's accessed system $AE$. This phase randomization will lead to the decoherence of system $AE$ and limit the information that can be gained by Eve.

\section*{Simulation}

To estimate the performance of our protocol, numerical simulation is given.
In this simulation, we use the polarization state of photon transmitted in optical fiber to realize the coding system.
On Bob's side, the home qubit is assumed to transmit in a round channel whose the efficiency is the same as the Bob-Alice-Bob channel for simplicity.
We use off-the-shelf experimental parameters to establish the simulation, e.g., optical fiber is of an attenuation of 0.20 $dB/km$, detection efficiency is $\eta_d=10\% $ and its dark count rate is $p_d=10^{-5}$. Besides, we consider a misalignment of detector as $d_e=1\%$,
$\eta_\rightarrow$ and $\eta_\leftarrow$ in the key rate generation formula just equal the transmission efficiency of the corresponding optical fiber $\eta$.
All the polarization error  corresponding to $p'_{01}$ comes from dark count of single photon detector,so $p'_{01}=\frac{\eta\eta_dd_e+(1-\eta\eta_d)p_d}{\eta\eta_d+2(1-\eta\eta_d)p_d}$.
For the error rate between Alice and Bob, it only comes from dark count as $e=\frac{(1-\eta^{2})\eta^{2}\eta_{d}p_{d}}{\eta^{4}\eta^{2}_{d}+2(1-\eta^{2})\eta^{2}\eta_{d}p_{d}}$.
The overall secure-key generation rate is thus $R=\eta^{4}(1-\frac{H(p'_{01})+H(p'_{10})}{2\eta}-H(e))$ per trial.
The simulation is shown in Fig.~\ref{Fig:simulation1}. This result show that the modified PP protocol can distribute secure-key bits for distant peers around $50$ km.

\section*{Discussion and Conclusion} \label{sec-conclusion}
Two-way deterministic QKD protocols, including PP protocol, do not require basis choices. Thus every encoded bit is used for final key generation. Besides, it has been proved some two-way deterministic QKD protocols are secure against detector-side-channel attacks on the backward channel\cite{Cai2013}. These advantages make such type protocols potentially useful.

The security of PP protocol when channel loss and noise are presented has been an open problem for a long time. To overcome this problem, we add a simple and experimentally feasible modification to the original PP protocol. Quite interestingly, our modification only leads to a trivial overall to the system of Alice and Bob, but can introduce a phase randomization for Eve's system. With the effects of this phase randomization, we prove the security of this modified PP protocol when the noisy and lossy channel is taken into account. Simulation results show that our protocol is practical. And we also hope that our modification on PP protocol can shed lights on other two-way QKD protocols.

\section*{Acknowledgments}
This work was supported by the National Basic Research Program of China (Grants No. 2011CBA00200 and No. 2011CB921200), National Natural Science Foundation of China (Grants No. 61101137 and No. 61201239).

\section*{Author contributions}

For this publication, Z.Y. proposed the protocol. Y.H. did the analysis. H.L. wrote the main manuscript and S.W. prepared figures 1-2. W.C., Z.H. and G.G.  provided essential comments to the manuscript.
All authors reviewed the manuscript.
The first two authors contributed equally in this letter.

\section*{Additional information}
Competing financial interests: The authors declare no competing financial interests.
\\

\newpage

Figure-1  Coding system of modified PP protocol. There are two modes: message mode(solid line) and control mode(dash line) in the operations of Alice and Bob. In message mode, if Alice wants to encode bit $0$, she will randomly perform $I_0$ or $I_1$. If Alice wants to encode bit $1$, she will randomly perform $Y_0$ or $Y_1$. The operations are defined as $I_{0}\{\ket{v},\ket{0},\ket{1}\}=\{\ket{v},\ket{0},\ket{1}\}$, $I_{1}\{\ket{v},\ket{0},\ket{1}\}=\{\ket{v},-\ket{0},-\ket{1}\}$,
$Y_{0}\{\ket{v},\ket{0},\ket{1}\}=\{\ket{v},\ket{0},-\ket{1}\}$, $Y_{1}\{\ket{v},\ket{0},\ket{1}\}=\{\ket{v},-\ket{0},\ket{1}\}$. BSM is for Bell-states measurement, and
 Z represents the measurement defined by projectors $\{|v\rangle\langle v|,|0\rangle\langle 0|, |1\rangle\langle 1|\}$.\\

Figure-2  Simulation: Secure-key rate $Lg(R)$ vs channel distance L (km) from Bob to Alice. We set $\eta_d=0.1,d=10^{-5}$ per pulse. The detector error rate is $1\%$.\\

\begin{figure}[hbt]
\centering
\includegraphics[width=7.5cm]{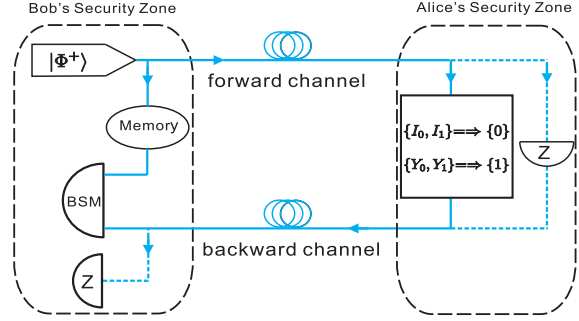}
\caption{Coding system of modified PP protocol. There are two modes: message mode(solid line) and control mode(dash line) in the operations of Alice and Bob. In message mode, if Alice wants to encode bit $0$, she will randomly perform $I_0$ or $I_1$. If Alice wants to encode bit $1$, she will randomly perform $Y_0$ or $Y_1$. The operations are defined as $I_{0}\{\ket{v},\ket{0},\ket{1}\}=\{\ket{v},\ket{0},\ket{1}\}$, $I_{1}\{\ket{v},\ket{0},\ket{1}\}=\{\ket{v},-\ket{0},-\ket{1}\}$,
$Y_{0}\{\ket{v},\ket{0},\ket{1}\}=\{\ket{v},\ket{0},-\ket{1}\}$, $Y_{1}\{\ket{v},\ket{0},\ket{1}\}=\{\ket{v},-\ket{0},\ket{1}\}$. BSM is for Bell-states measurement, and
 Z represents the measurement defined by projectors $\{|v\rangle\langle v|,|0\rangle\langle 0|, |1\rangle\langle 1|\}$.}
\label{Fig:codingsystem}
\end{figure}

\begin{figure}[hbt]
\centering
\includegraphics[width=7.5cm]{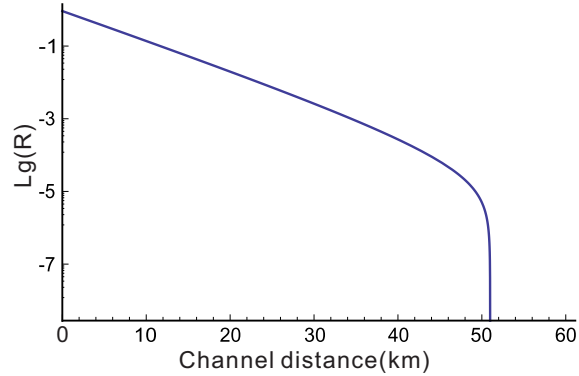}
\caption{Simulation: Secure-key rate $Lg(R)$ vs channel distance L (km) from Bob to Alice. We set $\eta_d=0.1,d=10^{-5}$ per pulse. The detector error rate is $1\%$. }
\label{Fig:simulation1}
\end{figure}

\end{document}